УДК 621.3.049.77::621.3.082.61::536.413.2

Л. С. Синев, В. Т. Рябов

# СОГЛАСОВАНИЕ КОЭФФИЦИЕНТОВ ТЕРМИЧЕСКОГО РАСШИРЕНИЯ ПРИ ЭЛЕКТРОСТАТИЧЕСКОМ СОЕДИНЕНИИ КРЕМНИЯ СО СТЕКЛОМ

Рассмотрены напряжения, возникающие в деталях из стекла и кремния после их соединения электростатическим методом из-за разницы их коэффициентов теплового линейного расширения. Представлена аналитическая модель, позволяющая графически определить оптимальную температуру проведения соединения. Даны рекомендации по выбору температуры проведения процесса соединения.

Ключевые слова: анодная посадка, электростатическое соединение, тепловое расширение, напряжение.

Leonid S. Sinev, Vladimir T. Ryabov

# COEFFICIENT OF THERMAL EXPANSION BALANCING FOR FIELD ASSISTED BONDING OF SILICON TO GLASS

Stresses caused by thermal expansion coefficients mismatch of anodically bonded glass and silicon samples are studied. An analytical model to determine graphically the optimum bonding temperature is presented. Recommendations on choosing bonding temperature are made.

Электростатическое соединение является одной из основных сборочных операций микросистемной техники. На фоне тенденции к миниатюризации измерительных и управляющих приборов и развития соответствующих технологий, всё большую важность приобретает снижение взаимного влияния материалов, контактирующих внутри одного прибора. В особенности это относится к напряжениям, возникающим после сборки кремниевых и стеклянных деталей микромеханических приборов электростатическим соединением.

При электростатическом соединении кремний соединяется со стеклом посредством приложения внешней разности потенциалов и одновременного нагрева до температур 200…450 °C [1, 2]. Кремниевая пластина помещается на стеклянную. Соединяемые поверхности нагревают. Отрицательный электрод источника высокого напряжения прикладывают к стеклянной пластине, а положительный электрод — к кремниевой. Значение приложенного напряжения составляет 600…1500 В.

Физически процесс соединения между стеклом и кремнием происходит следующим образом: при повышенной температуре, при наличии сильного электрического поля, положительные ионы натрия ($Na^+$) в стекле дрейфуют к отрицательному электроду на стеклянной пластине и нейтрализуются. Из-за их перемещения на границе с кремнием образуется избыточный отрицательный заряд, сформированный ионами кислорода. Происходит падение потенциала в области стекла, прилегающего к кремнию. Кислород из стекла соединяется с кремнием на границе раздела стекло-кремний. Таким образом, на границе появляется тонкий слой $SiO_2$, соединяющий кремниевую пластину и стеклянное основание. Изоляционные свойства этого слоя приводят к падению тока через соединение и процесс завершается.

В результате соединения образуются коэффициентные напряжения, то есть напряжения, возникающие вследствие разности значений коэффициентов теплового линейного расширения (КТЛР) стекла и кремния [3].

До начала нагрева стеклянная и кремниевая детали имеют одинаковые размеры, при нагреве детали расширяются неравномерно и при температуре соединения имеют отличающиеся размеры. После соединения детали, охлаждаясь, взаимно деформируются. При значении КТЛР стекла большем значения КТЛР кремния, снижение температуры вызывает напряжения растяжения в стекле и напряжения сжатия в кремнии. При значении КТЛР стекла меньшем значения КТЛР кремния, снижение температуры, наоборот, вызывает напряжения сжатия в стекле и напряжения растяжения в кремнии.

Целью данной работы является определение соотношения КТЛР стекла и кремния для обеспечения соединения с минимальными коэффициентными напряжениями. Нелинейная зависимость КТЛР соединяемых деталей от температуры не позволяет минимизировать коэффициентные напряжения путём подбора материалов с близкими средними КТЛР.

В рамках рассматриваемой модели считаем обе соединяемых детали сплошными, однородными, изотропными и непрерывными, представляющими по форме прямоугольные параллелепипеды. Используем допущение, что нагрев деталей равномерен и источник тепла расположен вне области соединения. Также считаем, что область соединения представляет собой плоскость. Изменения размеров рассматриваем в плоскости перпендикулярной плоскости соединения. Считаем, что деформации и напряжения в области соединения равны деформациям и напряжениям во всей детали. Влияние краевых эффектов и разницы в коэффициентах теплопроводности материалов исключаем из рассмотрения. Исходим из того, что толщина кремниевой детали примерно в 10 раз меньше каждого из двух других ее размеров, и в 5…10 раз меньше толщины стеклянной детали. Таким образом, изгиб деталей под действием возникающих деформаций пренебрежимо мал.

Распределение деформаций и напряжений по толщине соединяемых деталей будут учтены в последующих работах.

Истинным коэффициентом теплового линейного расширения $\alpha$, 1/°C, называется отношение изменения линейного размера тела $dl$, м, деленного на его начальный размер $l_0$, м, к малому изменению температуры $dT$, °C, вызвавшему изменение размера тела [4].

$$\alpha = \frac{1}{l_0} \cdot \frac{dl}{dT}$$

Поскольку детали соединены, то в каждый момент времени изменения длины должны быть равны.

$$dl_g = dl_{si},$$

где $dl_g$ — изменение длины стеклянной детали, м; $dl_{si}$ — изменение длины кремниевой детали, м.

Изменения длины, учитывая вышесказанное, определяются как тепловым расширением самой детали, так и влиянием присоединенной детали.

$$dl_{si} = \alpha_{si}(T) \cdot l_0 dT - \frac{dZ \cdot l_0}{E_{si} A_{si}}$$

$$dl_g = \alpha_g(T) \cdot l_0 dT + \frac{dZ \cdot l_0}{E_g A_g}$$

$$dZ = \frac{E_g A_g E_{si} A_{si}}{(E_{si} A_{si} + E_g A_g)} (\alpha_{si}(T) - \alpha_g(T)) dT$$

где $\alpha_g(T)$, $\alpha_{si}(T)$ — коэффициенты теплового линейного расширения стекла и кремния, соответственно, 1/°C;

$E_g$, $E_{si}$ — модули упругости первого рода стекла и кремния, соответственно, Па;

$A_g$, $A_{si}$ — площади сечений, расположенных поперек оси удлинения стекла и кремния, м²;

$dZ$ — изменение растягивающей силы в стекле из-за изменения температуры, Н.

$$\sigma_g(T) = \frac{\Delta Z}{A_g} = \frac{E_g E_{si} A_{si}}{(E_{si} A_{si} + E_g A_g)} \int_{T_c}^{T_0} (\alpha_{si}(T) - \alpha_g(T)) dT \qquad (1)$$

$$\sigma_{si}(T) = \frac{\Delta Z}{A_{si}} = \frac{E_g A_g E_{si}}{(E_{si} A_{si} + E_g A_g)} \int_{T_c}^{T_0} (\alpha_{si}(T) - \alpha_g(T)) dT \qquad (2)$$

где $\sigma_g$, $\sigma_{si}$ — напряжения при рабочей температуре, растягивающие в стекле и сжимающие в кремнии, соответственно, Па; $T_0$, $T_c$ — температуры рабочая и соединения, соответственно, °C.

Интеграл в формулах (1, 2) представляет собой разницу между относительными удлинениями кремния и стекла, если бы они не были соединены. Также видно, что соотношение напряжений в кремнии и стекле обратно пропорционально площадям их поперечных сечений.

Исходя из данных о КТЛР применяемых стекла и кремния, на основании вышеприведенных формул можно рассчитать значения коэффициентных напряжений при рабочей температуре $T_0$ в деталях соединенных при температуре $T_c$. Рассмотрим такой расчет на примере соединения кремния со стеклами марок ЛК-5 (Лыткаринский завод оптического стекла, Россия) и Corning 7740 (Corning, США).

Экспериментально полученная зависимость КТЛР кремния приведена в литературе [5]. Модуль Юнга рассматриваемого направления кремния [100] составляет 130,2 ГПа. Расчет зависимости КТЛР для стекол был проведен на основании данных, указанных производителями, исходя из предположения, что температурный ход значения коэффициента в области «нормального» состояния стекол, от минус 120 °C до температуры нижней границы зоны отжига, практически выражается линейным уравнением [4]. Для стекла ЛК5 [6]: модуль Юнга — 68,45 ГПа, КТЛР в диапазоне от минус 60 до плюс 20 °C составляет $33 \cdot 10^{-7}$ 1/°C, КТЛР в диапазоне 20…120°C — $35 \cdot 10^{-7}$ 1/°C. Для стекла Corning 7740 [7]: модуль Юнга — 62,76 ГПа, КТЛР в диапазоне 0…300°C — $32,5 \cdot 10^{-7}$ 1/°C, КТЛР в диапазоне 25…821°C — $35 \cdot 10^{-7}$ 1/°C. Расчеты числовых значений напряжений были проведены исходя из размеров деталей: стеклянной — 4 x 4 x 3 мм; кремниевой — 4 x 4 x 0,42 мм. Графики зависимостей представлены на рис. 1а.

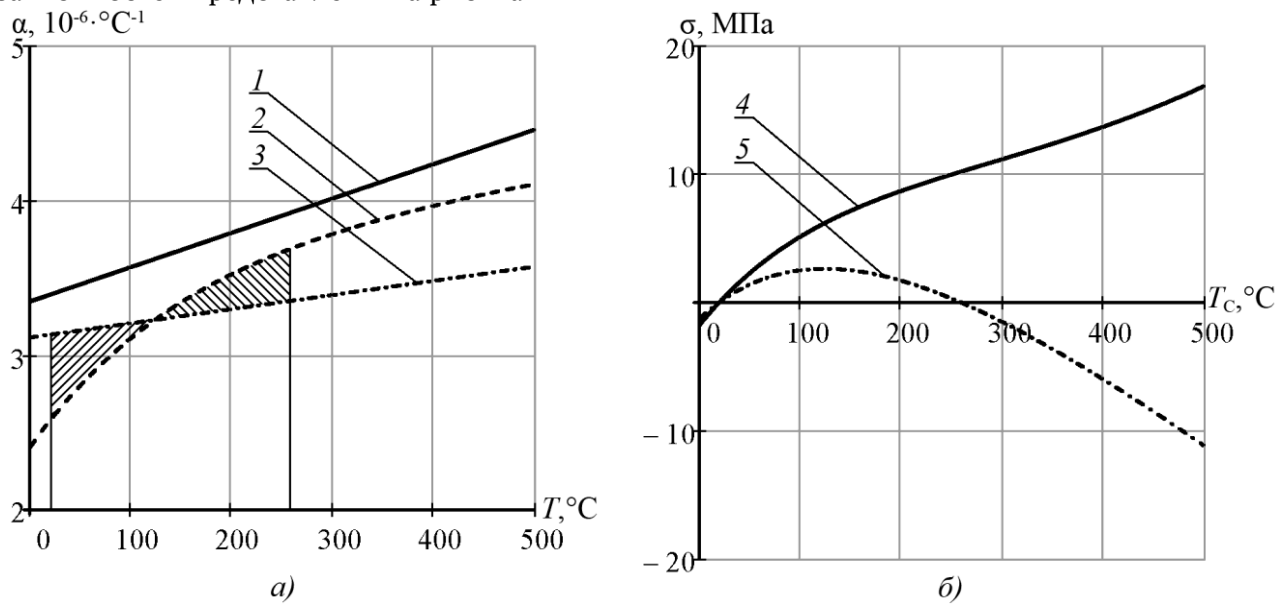

Рис. 1. Стекла ЛК-5, Corning 7740 и кремний. а) зависимость КТЛР стекла ЛК-5 (1), стекла Corning 7740 (3) и кремния (2) от температуры; б) коэффициентные напряжения в кремнии при соединении со стеклом ЛК-5 (4) и стеклом Corning 7740 (5) при рабочей температуре 20 °C в зависимости от температуры проведения соединения.

На основании зависимостей (1) и (2) можно определить напряжения в соединяемых деталях в зависимости от температуры. Если задаться рабочей температурой (температурой остывшего состояния), можно построить зависимость напряжений от температуры проведенного процесса электростатического соединения (см. рис. 1б). Таким образом, можно спланировать процесс соединения, обеспечивающий минимальные коэффициентные напряжения или же выдерживающий их в определенных пределах.

Задавшись температурой процесса соединения, и построив зависимость напряжений от рабочей температуры, можно оценить характер изменения коэффициентных напряжений в рабочем диапазоне температур получаемого изделия (см. рис. 2).

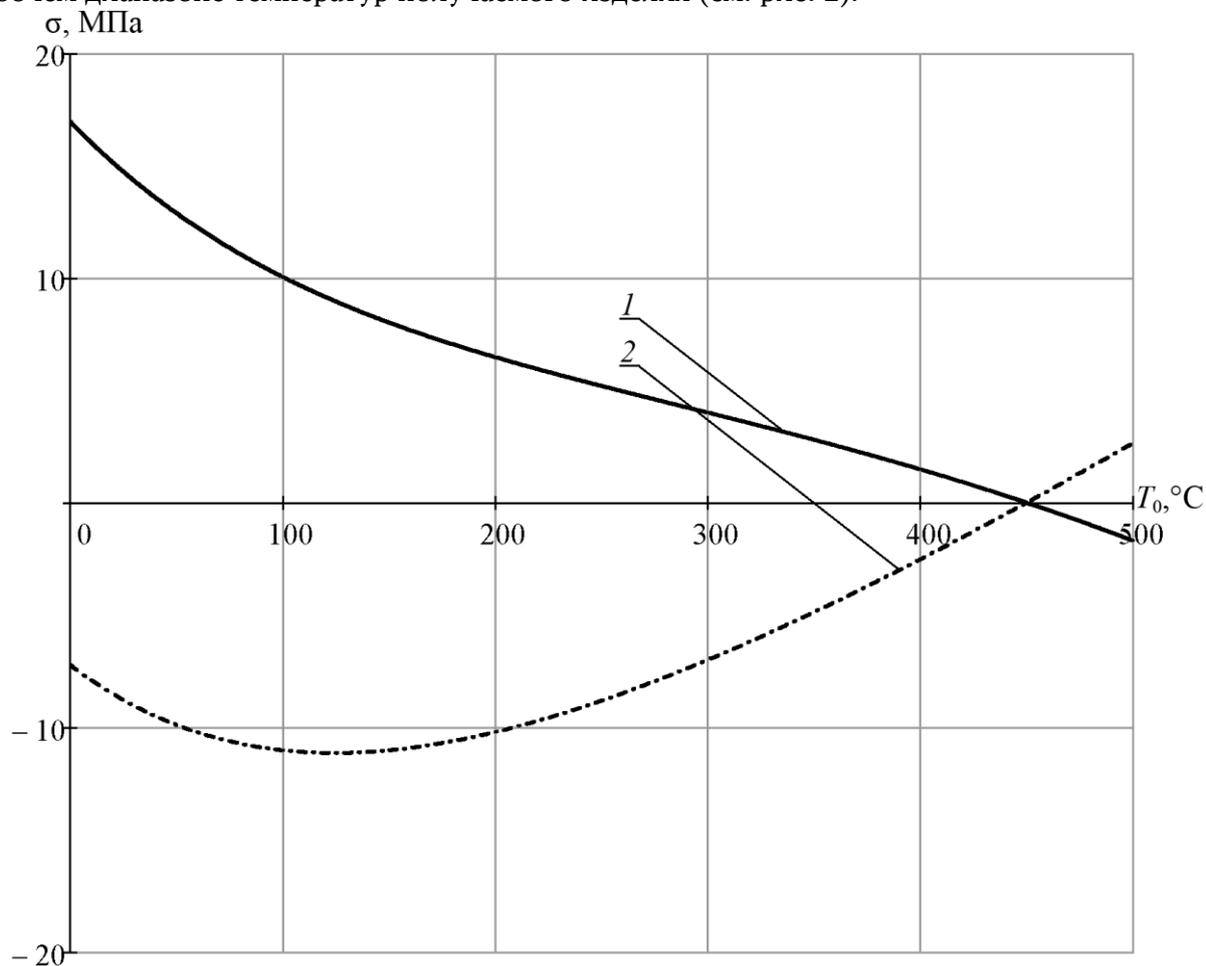

Рис. 2. Графики изменения коэффициентных напряжений в кремнии, соединенном со стеклом ЛК-5 (1) и стеклом Corning 7740 (2) в зависимости от рабочей температуры (при температуре проведения соединения 450 °C).

Проанализируем различные варианты соотношения истинных коэффициентов теплового линейного расширения стекла и кремния. Для каждого случая рассмотрим влияние разницы между этими коэффициентами на напряжения собранных изделий. Определим, как должны соотноситься КТЛР стекла и кремния для получения наименее напряженного соединения.

Первый случай — КТЛР стекла в рассматриваемой области температур всегда больше КТЛР кремния. В качестве примера возьмем рассмотренные ранее графики для стекла ЛК-5 и кремния (см. рис. 1а). В результате постоянно имеющейся разницы между коэффициентами теплового расширения, напряжения, вызванные этой разницей, будут постоянно увеличиваться. Увеличение значений напряжений будет тем больше, чем больше разница между КТЛР соединяемых материалов (см. рис. 2). В рассматриваемом случае коэффициентные напряжения будут тем больше, чем больше разница между температурой соединения и рабочей температурой прибора. Процесс электростатического соединения, обеспечивающий минимальные коэффициентные напряжения, следует проводить при наименьшей допустимой температуре (см. рис. 1б).

Следующий рассматриваемый случай — в одном диапазоне температур коэффициент теплового расширения стекла больше коэффициента кремния, а в другом диапазоне меньше. Таким образом, график зависимости КТЛР стекла от температуры пересекает график зависимости КТЛР кремния. Иллюстрацией данной ситуации послужат графики для стекла Corning 7740 и кремния (см. рис. 1а). При охлаждении, имеющаяся при соединении при повышенной температуре разница между КТЛР материалов будет уменьшаться и впоследствии сменит свой знак. Величина напряжений, вызванных этой разницей, будет увеличиваться до температуры, при которой коэффициенты соединяемых материалов будут равны. При дальнейшем снижении температуры величина напряжений будет уменьшаться до точки смены знака напряжений (их направленности). В этой точке коэффициентные напряжения будут равны нулю. Определить температуру проведения процесса электростатического соединения, обеспечивающую отсутствие напряжений при заданной рабочей температуре, можно графически. Области, отсеченные линиями температур и графиков КТЛР стекла и кремния по обе стороны от точки пересечения графиков КТЛР, должны быть равными (см. рис. 1а и 1б, для Corning 7740 и рабочей температуры 20 °C температура соединения, обеспечивающая отсутствие коэффициентных напряжений, составляет 260 °C).

Также возможна ситуация, когда в интервале между рабочей температурой изделия и предполагаемой температурой процесса электростатического соединения графики зависимостей КТЛР соединяемых материалов пересекаются дважды (см. рис. 3а).

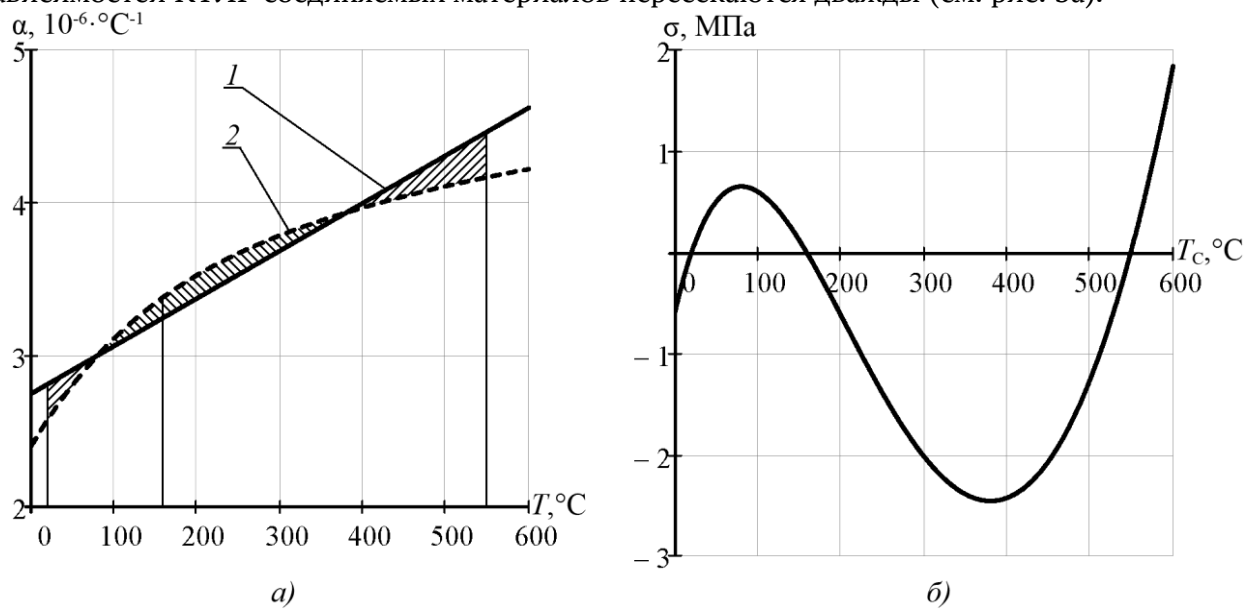

Рис. 3. График изменения КТЛР стекла пересекает график КТЛР кремния дважды.
а) зависимость КТЛР стекла (1) и кремния (2) от температуры; б) напряжения в кремнии при рабочей температуре 20 °C в зависимости от температуры проведения соединения.

Тенденции, описанные выше, имеют место и в этом случае. Аналогично предыдущему примеру, напряжения имеют экстремумы значений при температурах, соответствующих пересечений графиков коэффициентов тепловых расширений соединяемых материалов (см. рис. 3б). Таким же образом проводят графическое определение желательной температуры проведения процесса электростатического соединения. Площади, отсеченные границами температур и графиками КТЛР по одну сторону от какого-либо графика КТЛР, должны быть равны суммарной площади, отсеченной по другую сторону того же графика КТЛР.

В случае, когда графики КТЛР стекла и кремния касаются в одной точке (см. рис. 4а), коэффициентные напряжения будут увеличиваться, как и в первом из описанных выше случаев. Однако следует отметить, что вблизи температуры, при которой коэффициенты теплового линейного расширения материалов будут равны, значения напряжений будут изменяться довольно мало (см. рис. 4б).

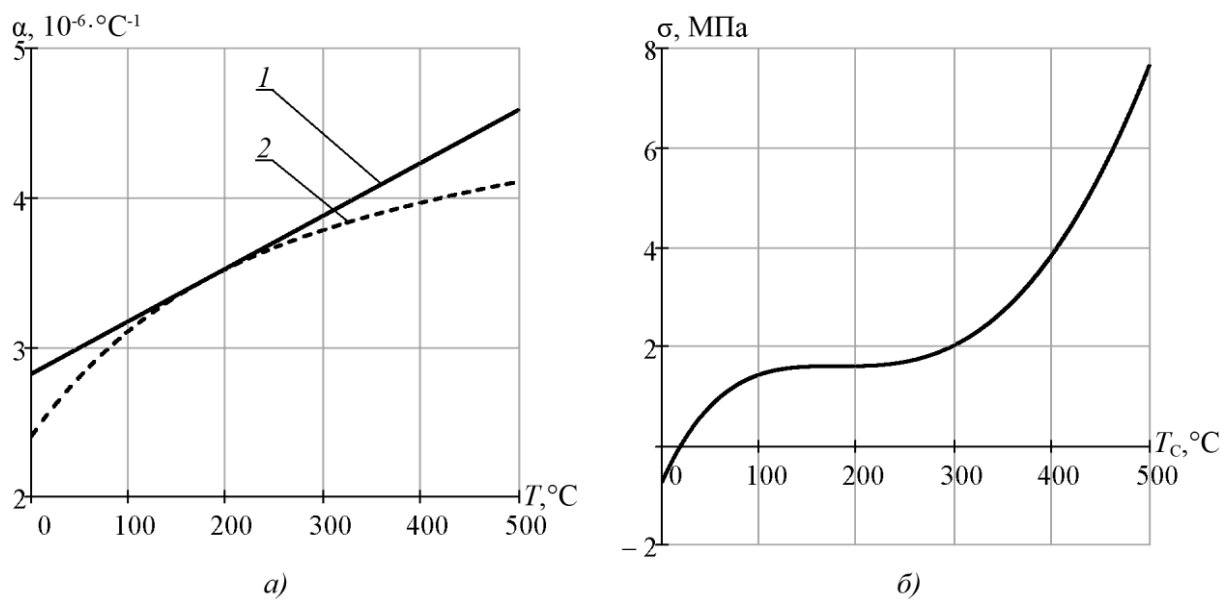

Рис. 4. График изменения КТЛР стекла касается графика КТЛР кремния.
а) зависимость КТЛР стекла (1) и кремния (2) от температуры; б) напряжения в кремнии при рабочей температуре 20 °C в зависимости от температуры проведения соединения.

Таким образом, если диапазон рабочих температур прибора будет находиться вблизи точки равенства КТЛР, коэффициентные напряжения будут слабо зависеть от изменения температуры прибора.

Для получения наименее напряженного соединения кремния со стеклом в общем случае, необходимо:
1) учитывать зависимость КТЛР стекла и кремния от температуры и обеспечить минимальную накапливаемую разницу между ними в процессе охлаждения (см. рис. 1а и рис. 3а):
   – анодную посадку для стекла 7740 фирмы Corning проводить при температуре 250...300 °C;
   – анодную посадку для стекла ЛК-5 проводить при минимально допустимой для успешного проведения процесса температуре;
2) провести исследования для определения химического состава стекла с КТЛР, максимально приближенным к зависимости КТЛР кремния от температуры.

**Список использованной литературы**